# Coexisting Ferromagnetic–antiferromagnetic Phase and Manipulation in Magnetic Topological Insulator MnBi$_4$Te$_7$

Jianfeng Guo[1+], Huan Wang[1+], Xueyun Wang[3], Shangzhi Gu[1], Shuo Mi[1], Shiyu Zhu[2], Jiawei Hu[2], Fei Pang[1], Wei Ji[1], Hongjun Gao[2], Tianlong Xia[1*] and Zhihai Cheng[1*]

[1]Beijing Key Laboratory of Optoelectronic Functional Materials & Micro-nano Devices, Department of Physics, Renmin University of China, Beijing 100872, China

[2]Institute of Physics, Chinese Academy of Sciences, China

[3]School of Aerospace Engineering, Beijing Institute of Technology, Beijing 100081, China

**Abstract:** Magnetic topological insulators (MTIs) have received a lot of attention due to the existence of various quantum phenomena such as quantum anomalous Hall effect (QAHE) and topological magnetoelectric effect, etc. The intrinsic superlattice-like layered MTIs, MnBi$_2$Te$_4$/(Bi$_2$Te$_3$)$_n$, have been extensively investigated mainly through the transport measurements, while the direct investigation of their superlattice-sensitive magnetic behaviors is relatively rare. Here, we report a microscopic real-space investigation of magnetic phase behaviors in MnBi$_4$Te$_7$ using cryogenic magnetic force microscopy (MFM). The intrinsic robust A-type antiferromagnetic (AFM), and emerged surface spin-flop (SSF), canted AFM (CAFM), ferromagnetic (FM)+CAFM, forced FM phases are sequentially visualized via the increased external magnetic field, in agreement with the metamagnetic behavior in the M-H curve. The temperature-dependent magnetic phase evolution behaviors are further investigated to obtain the complete H-T phase diagram of MnBi$_4$Te$_7$. The tentative local phase manipulation via the stray field of the magnetic tip is demonstrated by transforming the AFM to FM phase in its surface layers of MnBi$_4$Te$_7$. Our study the not only provide key real-space ingredients for understanding their complicated magnetic, electronic, and topological properties of these intrinsic MTIs, but also suggest new directions for manipulating spin textures and locally controlling their exotic properties.

+ These authors contributed equally to this work.
* To whom correspondence should be addressed: zhihaicheng@ruc.edu.cn tlxia@ruc.edu.cn

**Introduction**

Magnetic topological insulators (MTIs) are of great interest due to the existence of various exotic phenomena, such as quantum anomalous Hall effect (QAHE), topological magnetoelectric effect, Majorana modes, magnetic Dirac and Weyl semimetals, axion insulators, topological superconductors, etc.[1-5] Magnetism plays an extremely important role in the realization of various topological phenomena, due to its breaking of time-reversal symmetry. The QAHE previously observed in magnetically doped TI thin films can only be realized at extreme low temperatures due to the inhomogeneity of the sample.[6-8] However, the advent of stoichiometric materials with a spontaneously broken time-reversal symmetry provides a new platform to study various topological states.[4, 9, 10]

Recently, $MnBi_2Te_4$ with abundant intrinsic magnetic phases was shown to be the first antiferromagnetic (AFM) TI.[9, 11-24] The intrinsic magnetic phases are crucial for the realization of novel topological phases and transitions. QAHE and axion insulator phase exist in odd and even numbers of septuple layers (SLs) in the intrinsic AFM phase, respectively.[13, 19, 21] The Chen insulator phase transition exists in the intrinsically canted AFM (CAFM) phase.[20, 24] The realized temperature of QAHE can be effectively increased in the forced ferromagnetic (FM) phase.[19] Meanwhile, the manipulation of the microscopic magnetic phase can achieve the modulation of the quantum states.[25-28] However, due to the strong interlayer coupling $J$, the modulation of different magnetic phases of $MnBi_2Te_4$ requires a strong magnetic field. Also, the manipulation of the microscopic magnetic phases is difficult to achieve. Interestingly, because the $MnBi_2Te_4$ (MBT) layers are stacked by van der Waals interactions, a series of superlattice-like layered $MnBi_2Te_4/(Bi_2Te_3)_n$ (n=1, 2, 3...) can be formed by inserting $Bi_2Te_3$ (BT) layers, which can effectively reduce the interlayer coupling $J$.[29-32] The reduction of the interlayer coupling $J$ not only drastically reduces the magnetic field used for magnetic phase modulation, but also provides the possibility of manipulating the microscopic magnetic phases. However, it also makes the magnetic behavior of the sample more elusive.[29-43] Therefore, Real-space visualization of magnetic phase behaviors is essential to study the topological properties of the sample.

**In this work,** we report the real-space visualization of different magnetic phases in MnBi$_4$Te$_7$ single crystal using cryogenic magnetic force microscopy (MFM). With increasing magnetic field, MnBi$_4$Te$_7$ single crystal can be divided into five magnetic phases, namely, the intrinsic robust A-type AFM phase, the surface spin-flop (SSF) phase, the CAFM phase, the FM+CAFM phase, and the forced FM phase. Among them, the novel FM+CAFM coexistence phase induces the metamagnetic behavior of MnBi$_4$Te$_7$ single crystal. The existence of multiple magnetic phases provides an excellent platform on the realization of different topological states. By further temperature-dependent studies, we obtained the complete H-T phase diagram of MnBi$_4$Te$_7$ single crystal. Finally, the manipulation of the microscopic magnetic structure becomes possible due to the reduction of the interlayer coupling *J*. Here, we successfully achieved the precise manipulation of the surface domains of MnBi$_4$Te$_7$ single crystal from the AFM to FM phase by using the stray field of the MFM tip. Our results not only provide a new venue for exploring and realizing various topological states, but also shed new light for advancing the application of AFM spintronic technologies.

## Results and Discussion

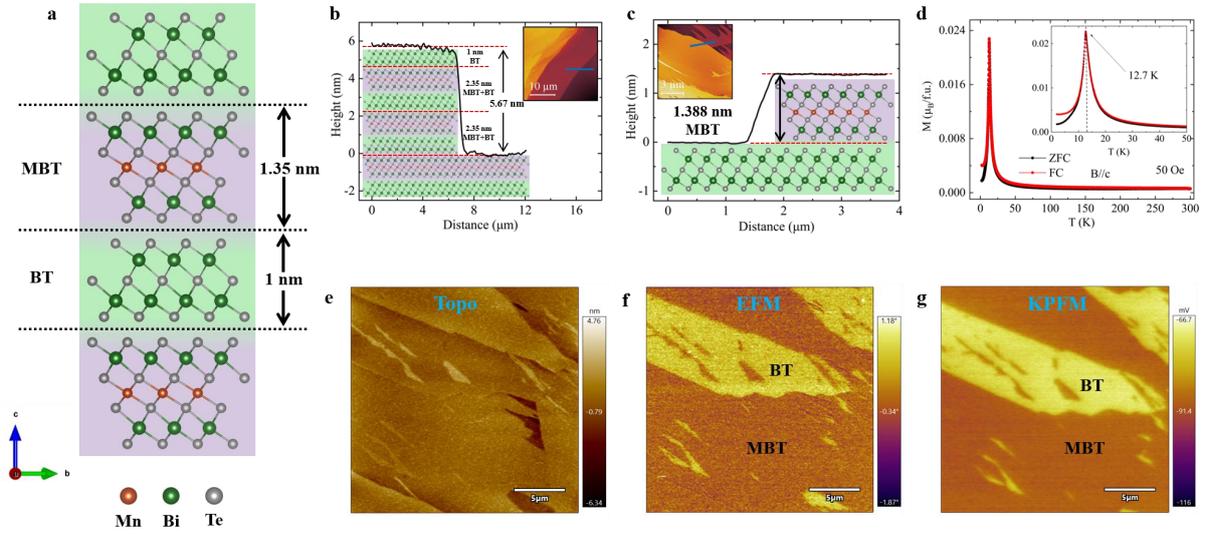

**Fig. 1. Crystal structure and AFM characterizations of MnBi$_4$Te$_7$ single crystal. (a)** Schematic illustration of the crystal structure of MnBi$_4$Te$_7$. Layered crystal structures, including the Bi$_2$Te$_3$ quintuple layer (BT, ~1.0 nm) and MnBi$_2$Te$_4$ septuple layer (MBT, ~1.35 nm). **(b,c)** Line profile of the AFM topography image (insets) with respective section graphs of cleaved MnBi$_4$Te$_7$ surfaces. **(d)** Temperature dependence of zero-field-cooled (ZFC) and field-cooled (FC) magnetic susceptibility for B//c. **(e)** Topography image of MnBi$_4$Te$_7$. **(f,g)** The corresponding EFM image and KPFM image of (e). The surface terminated-BT and -MBT layer are determined, respectively.

The Mn$_4$BiTe$_7$ single crystal is a superlattice-like layered material formed by alternating Bi$_2$Te$_3$ (BT, ~1.0 nm) and MnBi$_2$Te$_4$ (MBT, ~1.35 nm) layers, as shown in Fig. 1(a). The single-crystal x-ray diffraction (XRD) pattern reveals the (00l) crystalline surface of MnBi$_4$Te$_7$ (Fig. S1), showing no additional impurity phases. As a superlattice-like layered material of MnBi$_4$Te$_7$, there are two cleaving ways between the alternating MBT and BT layers, resulting in either MBT- or BT-terminated surfaces. After the cleaving process, two typical surface regions were first characterized by AFM, as shown by the topography images and line profiles of Figs. 1(b) and (c). The observed terrace heights of 5.67 nm and 1.39 nm, which are consistent with the BT+2MBT+2BT and MBT layer thicknesses, corresponding to two different termination surfaces, respectively. Fig. 1(d) shows temperature dependence of zero-field-cooled (ZFC) and field-cooled (FC) magnetic susceptibility for B//c. Clearly, a paramagnetic (PM) to AFM transition was observed at approximately T$_N$ ~12.7 K, consistent with previous reports.[29-31, 33-39]

To direct determine the surface terminations, bedside the surface topography image of Fig. 1(e), the electrostatic force microscope (EFM)[44, 45] and Kelvin probe force microscope (KPFM) measurements were performed on the same region of Fig. 1(e), as shown in Figs. 1(f) and (g). The EFM image clearly shows two distinct regions with different contrast, in consistent with two kinds of the BT- and MBT-terminated surfaces. Based on the quantitative KPFM measurements, the BT-terminated surfaces are further directly determined by their higher surface potential (~16.5 mV) than that of the MBT-terminated surfaces (Fig. S2). It is worth noting that the signal intensities of EFM phase and KPFM potential are only related to the type of termination surface instead of the morphological thickness. The different long-range electrostatic forces from the BT- and MBT-terminated surfaces could affect the obtained MFM signals of $MnBi_4Te_7$ below the Néel ordering temperature ($T_N$~12.7 K), which should be carefully discussed and discriminated. When the applied external magnetic fields completely polarized the $MnBi_4Te_7$ crystal into the force FM phase, the magnetic contrast completely disappears, and the obtained signal contrast in MFM images comes exclusively from the different long-range electrostatic force of BT- and MBT-terminated surfaces (Fig. S3).

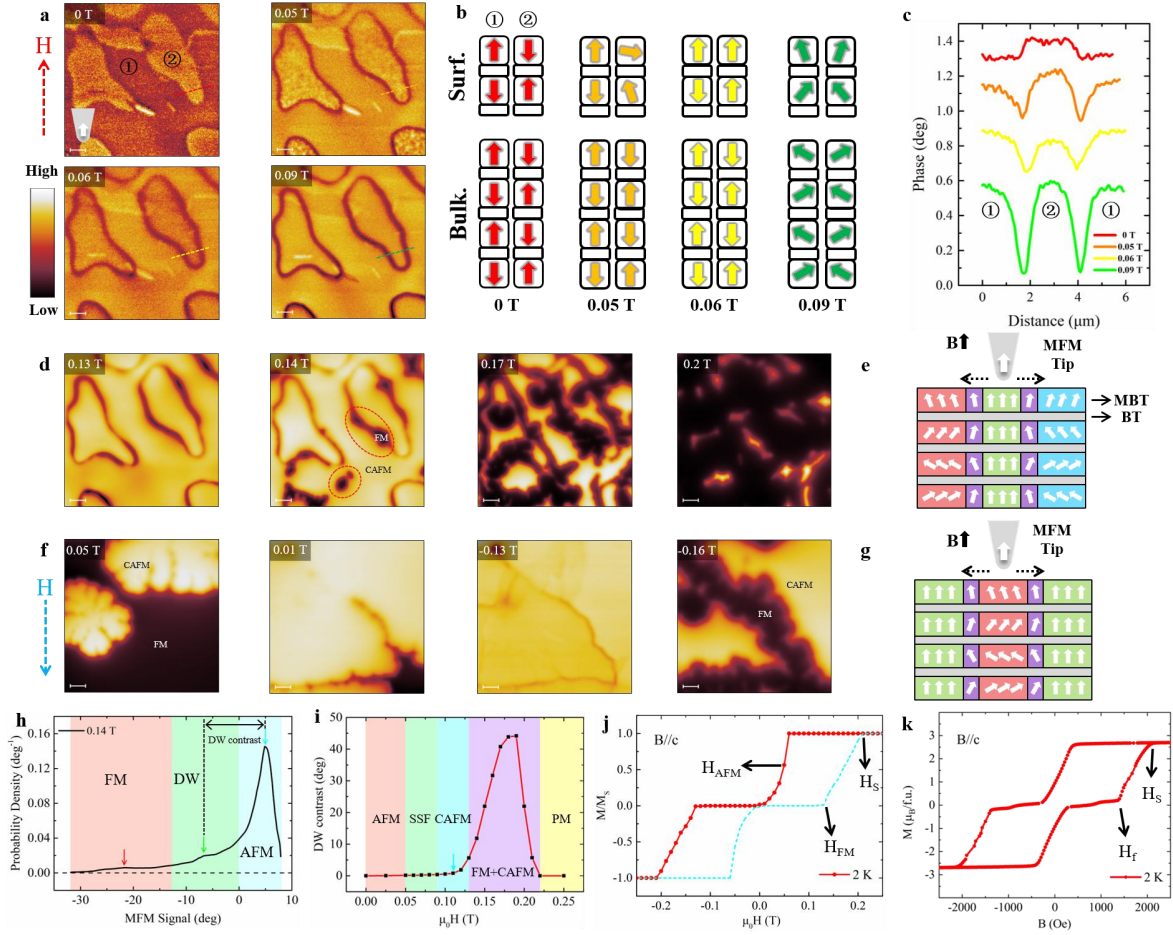

**Fig. 2. MFM characterization of the H-dependent magnetic phases in MnBi$_4$Te$_7$.** (a) MFM images of MnBi$_4$Te$_7$ taken at 2 K with increasing magnetic fields of 0, 0.05, 0.06, 0.09 T. The magnetic field is parallel to *c* and is taken to be positive in the upward direction. (b) Line profiles of the domain along the dash line in (a), showing a significant reversal of domain contrast. (c) Schematic illustration of the surface spin-flop (SSF) transition for (b). (d) MFM images with increasing magnetic fields of 0.13, 0.14, 0.17, 0.2 T. (e) Schematic diagram for the formation of FM phase at the DWs of AFM phase. (f) MFM images with decreasing magnetic fields of 0.05, 0.01, -0.13, -0.16 T. (g) Schematic diagram for the formation of AFM phase within the FM phase. (h) Histogram of the MFM signal at 0.14 T. (i) H-dependence of DW contrast. (j) The M-H hysteresis loop deduced from the H-dependent MFM images. (k) Magnetic hysteresis loop for B//c. Scale bar: 2 μm.

There are five different magnetic phases realized in the MnBi$_4$Te$_7$ under the variable external magnetic fields, namely, intrinsic robust A-type AFM phase (0 T-0.05 T), SSF phase (0.05 T-0.09 T), CAFM phase (0.09 T-0.13 T), FM+CAFM phase (0.13 T-0.22 T), and forced FM phase ( > 0.22 T) (Fig. S7). Despite the insertion of the BT layer, the MnBi$_4$Te$_7$ crystal still exhibits the same robust A-type AFM order as MnBi$_2$Te$_4$.[23]

Fig. 2(a) show typical MFM images obtained with increasing magnetic fields from 0 to 0.09 T. For an intrinsic A-type AFM with uniaxial anisotropy, there are only two possible AFM domain states, up-down-up-down (↑↓↑↓) and down-up-down-up (↓↑↓↑).[22, 23] For the MFM image at 0 T, two kinds of AFM domains, labeled ① and ②, can be clearly resolved via the relative dark (and bright) contrast, indicating the attractive (and repulsive) interaction, i.e., the surface magnetic moment is parallel (and antiparallel) to the tip magnetic moment. Here, the tip magnetic moment is upward, the AFM domains ① and ② are determined to be the up-down-up-down (↑↓↑↓) and down-up-down-up (↓↑↓↑) domains, respectively (Fig. S6). The typical domain size is ~10 μm, so the chiral edge states at domain walls (DWs) have limited effect on the topological states. It is noteworthy that some regions in single AFM domains show opposite signal contrast, which is caused by the lacking of MBT-terminated surface layer (Fig. S4). This result indicates that the $MnBi_4Te_7$ surface is robust A-type order, which can be further demonstrated by the presence of the SSF transition.

With the increase of magnetic field, the domain ② first shows the surface-spin-flop (SSF) transition (0.05 T) due to the antiparallel topmost magnetic moment with the magnetic field, and finally the topmost magnetic moment of domain ② is same with that of domain ① (0.06 T). The signal contrast of domain ② is decreased in compare with domain ①, as shown in Fig. 2(b) (Fig. S6). With the further increase of magnetic field, the bulk-spin-flop (BSF) transition occurs in either domain ① and ②, and the CAFM phase emerges in the whole $MnBi_4Te_7$ crystal, as shown in Fig. 3(b). In the CAFM phase (0.09 T), the magnetic signal contrasts of domains ① and ② are identical due to their same out-of-plane spin configurations. The results of the robust A-type order observed from our MFM images can exclude the surface relaxation hypothesis, so the reason for the presence of a gapless Dirac cone observed in angle-resolved photoemission spectroscopy (ARPES) remains to be resolved.[41]

The AFM domains ① and ② are separated by the antiphase DWs with a net out-of-plane moment and high magnetic susceptibility due to their spin configuration (Fig. S4).[22, 46] A more significant magnetization occurs in these DWs with stronger net moment than the AFM domains, as shown in Fig. 2(c). These magnetized DWs could be the nucleation center of the newly emerged FM phase. Fig. 2(d) shows typical MFM images obtained with increasing magnetic fields from 0.09 to 0.22 T (Fig. S7). When the magnetic field reaches the critical $H_{FM}$ (~0.13T) (Fig. S6), the FM phase starts to form and grown from the DWs, then appears and grow within the AFM domains, which are schematically shown in Fig. 2(e) and Fig. S5. When the magnetic field further increases to the critical $H_S$ (~0.22T), the FM phase area gradually increases, and the AFM phase disappears completely, then the whole crystal enters into the force FM phase.

Fig. 2(f) shows typical MFM images taken with decreasing magnetic fields from 0.22 T to -0.21 T (Fig. S8). Until the magnetic field is decreased to $H_{AFM}$ (~0.05T), the AFM phase emerge and grow within the FM phase, as schematically shown in Fig. 2(g), and in consistent with its large hysteresis nature.[29-34, 36-39] The growth of AFM phase region also shows a significant relaxation (Fig. S9). The FM phase region gradually disappears and reappears at the field of -0.13 T. Finally, the whole crystal enters into the opposite force FM phase.

Fig. 2(h) shows a histogram of the MFM signal at 0.14 T, in which the FM, DW and AFM phase regions are determined and marked, respectively. The DW contrast relative with AFM phase can be further conceived as the indicator to separate the five magnetic phases in combination with the H-dependent MFM images, as shown in Fig. 2(i). The hysteresis loop can also be deduced from the obtained MFM images, as shown in Fig. 2(j), in which the normalized magnetization ($M/M_S$) is reasonably estimated as the proportion of FM phase in the MFM images. Fig. 2(k) shows the M-H loop of MnBi$_4$Te$_7$ obtained via the magnetic transport measurements, in good agreement with the hysteresis loop of Fig. 2(j).

In previous transport study, the $H_f$ of ~0.135 T has been generally ascribed as the critical field where the MnBi$_4$Te$_7$ single crystal enters the CAFM phase.[29, 30, 36-38] Based on our real-space MFM measurements, the delicate surface spin-flop and bulk spin-flop transition sequentially occurs at ~0.05 T and 0.09 T, respectively. While the $H_f$ of ~0.135 T is confirmed as the critical field of $H_{FM}$ at here, where the MnBi$_4$Te$_7$ single crystal enters the FM+CAFM phase due to the locally emergence of FM phase.

Most uniaxial AFMs have much higher exchange energy than anisotropic energy, e.g., MnBi$_2$Te$_4$, so their magnetic properties are dominated by exchange energy. For MnBi$_4$Te$_7$, the presence of BT intercalation leads to a significant reduction of $J/J_F$ ($J$, interlayer exchange coupling; $J_F$, intralayer exchange coupling; MnBi$_2$Te$_4$, $J/J_F$≈-0.92; MnBi$_4$Te$_7$, $J/J_F$≈-0.04),[32] which eventually leads to delicate energy competition, and then the metamagnetic behavior observed at here, especially the locally emerged FM phase region. It is noteworthy that the uniaxial perpendicular anisotropy $K$ also play important roles in controlling the metamagnetic behavior of $MnBi_4Te_7$, which is a temperature-sensitive quantity in comparison with its intrinsic interlayer and intralayer exchange coupling. To further investigate its temperature-dependent magnetic phases, variable temperature H-dependence MFM measurements were further performed to complete the H-T phase diagram of $MnBi_4Te_7$.

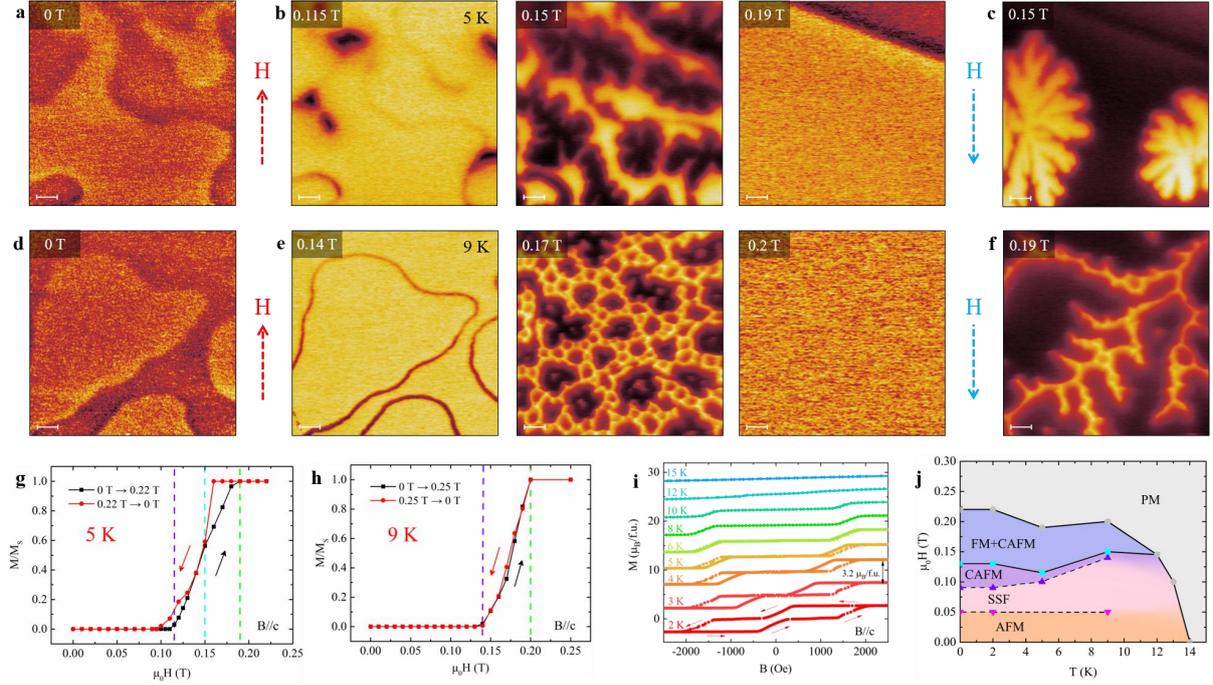

**Fig. 3. Variable temperature H-dependence MFM and transport measurements. (a,d)** MFM images taken at 0 T. **(b,c)** The corresponding MFM images taken at 5 K with increasing magnetic fields of 0.115, 0.15, 0.19 T (b) and decreasing magnetic fields of 0.15 T (c). **(e,f)** The corresponding MFM images taken at 9 K with increasing magnetic fields of 0.14, 0.17, 0.2 T (e) and decreasing magnetic fields of 0.19 T (f). **(g,h)** The M-H hysteresis loop deduced from MFM images at 5 K and 9 K, respectively. The purple, blue and green dashed lines mark $H_{FM}$, $H_{AFM}$ and $H_S$, respectively. **(i)** Magnetic hysteresis loop at variable temperatures for B//c. **(j)** H-T phase diagram showing A-type AFM phase (orange), SSF phase (pink), CAFM phase (purple), FM+CAFM phase (blue) and forced FM phase (gray). Scale bar: 2 μm.

Figs. 3(b) and (c) shows representative MFM images taken with various magnetic field at 5 K (Fig. S10). The crystal still can sequentially enter the CAFM, FM+CAFM and forced FM phases, while the AFM region represent a branching-like shape in this process (0.15 T), which is associated with the weakening of the uniaxial perpendicular anisotropy $K$. After the saturated magnetization, the AFM phase begins to reappear at ~0.15 T and shows a snowflake-like shape, indicating a significant reduction in hysteresis loop compared with that at 2 K. The hysteresis loop at 5 K is further deduced and plotted in Fig. 3(g).

Figs. 3(e) and (f) shows representative MFM images taken with various magnetic field at 9 K (Fig. S11 and S12), and the deduced hysteresis loop is shown in Fig. 3(h). In comparison with that at 5 K, there are two points worth emphasizing. First, after the appearance of FM phase, the AFM phase region shows an irregular honeycomb structure, indicating the high-density nucleation center of FM phase at 9 K. Second, although the hysteresis disappears completely, i.e., $H_{AFM}$ is equal to $H_S$, the local FM phase is still present, showing a soft magnetic state (Fig. S13). The similar hysteresis disappearance has been recently reported in MnBi$_4$Te$_7$,[29, 30, 36, 37] which have been ascribed to the occurrence of a general spin-flip transition. While the direct real-space MFM measurements confirm a persistent spin-flop transition to the CAFM phase even without discernible hysteresis at 9 K. The above two features could be readily elucidated by the Stoner-Wohlfarth model via the decreased $K/J$ ratio at higher temperatures.[32] When the temperature further increases to 12 K, the $K/J$ ratio has been decreased to a level that is insufficient to support the appearance of the local FM phase (Fig. S14). It is noteworthy that the remanent magnetic state can be fully magnetized when the temperature is low enough according to the relationship between temperature and hysteresis. Therefore, the high-resolution band structure of MnBi$_4$Te$_7$ in different magnetic phases (intrinsic AFM phase and the forced FM phase) may be able to be observed by ARPES.

Fig. 3(i) shows the temperate-dependent M-H loop of MnBi$_4$Te$_7$ single crystal. As the temperature increases, the hysteresis gradually decreases and completely disappears at ~8 K, which is consistent with the MFM results. Based on the above experimental results, the H-T phase diagram of the MnBi$_4$Te$_7$ was obtained, as shown in Fig. 3(j). It is noting that AFM phase can persist up to 14 K (>$T_N$ ~12.7 K) according to our MFM measurements (Fig. S14), which may be related to the surface layer effect.

The FM phase is not only essential for the realization of the QAHE, but also can effectively increase the observed temperature of the QAHE. Recently, Y. Tokura et. al., have realized the delicate control of chiral edge states in QAHE of magnetic-doping topological insulator via the designed and manipulation of magnetic domains with the MFM tip.[27] Meanwhile, the manipulability of the surface of MnBi$_4$Te$_7$ single crystal can be demonstrated by the presence of SSF transitions. Therefore, we would like to realize a locally controllable ferromagnetic phase in the antiferromagnetic phase. Based on the H-T phase diagram, the upper critical point of the CAFM phase (0.12 T) satisfies our experimental requirements.

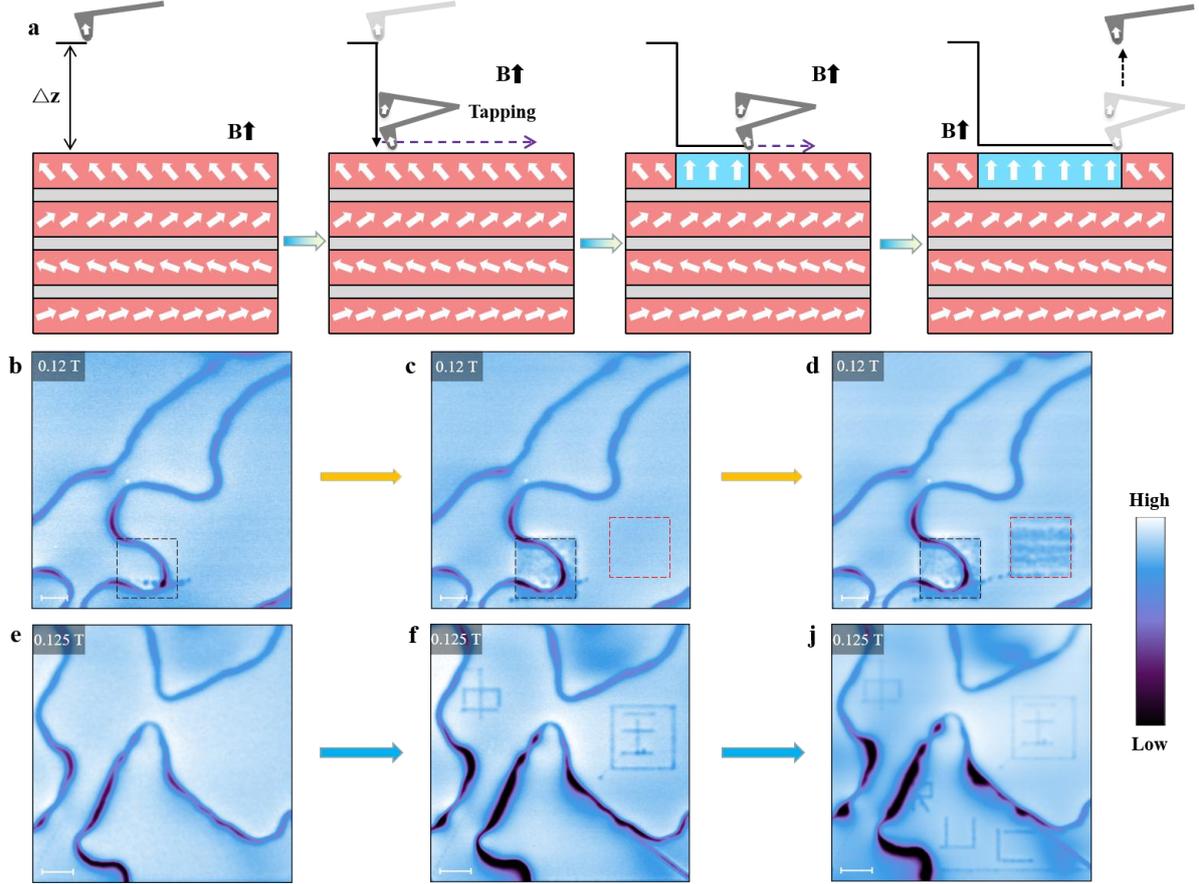

**Fig. 4. Local magnetic manipulation from AFM to FM phase in MnBi$_4$Te$_7$. (a)** Schematic diagram of the manipulation process and phase transition in the surface layers of MnBi$_4$Te$_7$. **(b-d)** MFM images taken during the scanning manipulation with the magnetic fields of 0.12 T. The black and red dashed squares represent the first and second scanning manipulation regions, respectively. **(e-j)** MFM images taken during the lithography manipulation with the magnetic fields of 0.125 T. Scale bar: (b-d) 2 nm; (e-j) 3 nm.

The tentative local phase manipulation via the stray field of the magnetic tip is demonstrated by transforming the AFM to FM phase in its surface layers of MnBi$_4$Te$_7$. Fig. 4(a) shows a schematic illustration of our manipulation procedure via the MFM tip. First, the original normal MFM image of the sample in the CAFM phase was obtained with the external magnetic field determined based on the H-T phase diagram. Then, the height of MFM tip is decreased to locally transform the surface layer into FM phases from CAFM via the stray magnetic field from the MFM tip. The designed FM domain regions or complicated patterns could be realized by controlled continues manipulation or lithographic "writing". Finally, the MFM image of the sample was obtained again to check the obtained magnetic domains after the previous manipulation.

Figs. 4(b-d) show the three sequential MFM images before, during and after the continues manipulation across the AFM-DW and within the single CAFM region. The manipulated dashed-box regions represent relative dark contrast (attractive interaction with the MFM tip) in contrast with that of CAFM, indicating that the successful realization of FM phase. It is noting that the dark contrast of manipulated region is weaker than that of DWs and nucleated FM region, which is in consistent with our proposed idea of surface-layered manipulation. More designed and complicated FM patterns can be "painted" via the lithographic manipulation, as shown in Figs. 4(e-f), demonstrating its high accuracy and prominent applications in magnetic topological electronics. Notably, the above magnetic manipulation in $MnBi_2Te_4$ was unprocurable in our experiments due to its stronger interlayer coupling than the stray field of tip.

The manipulation capability and accuracy are related to the distribution of tip stray field. According to previous reports,[27, 47] the magnetic tip can provide stray fields up to several thousand Gauss, but the decay of the stray field with distance is significant. The effective manipulation range is several tens of nanometers in the *xy*-plane and a several nanometers in the *z*-direction. By a precise magnetic manipulation, it is even possible to realized adjustable vertical AFM/FM heterostructure in the thin films of topological insulators, which could possess more exotic quantum transport properties. Due to the existence of SSF phase, $MnBi_4Te_7$ thin films have a broader research space and value, while the effect of finite size effect on SSF phase also needs to be studied.

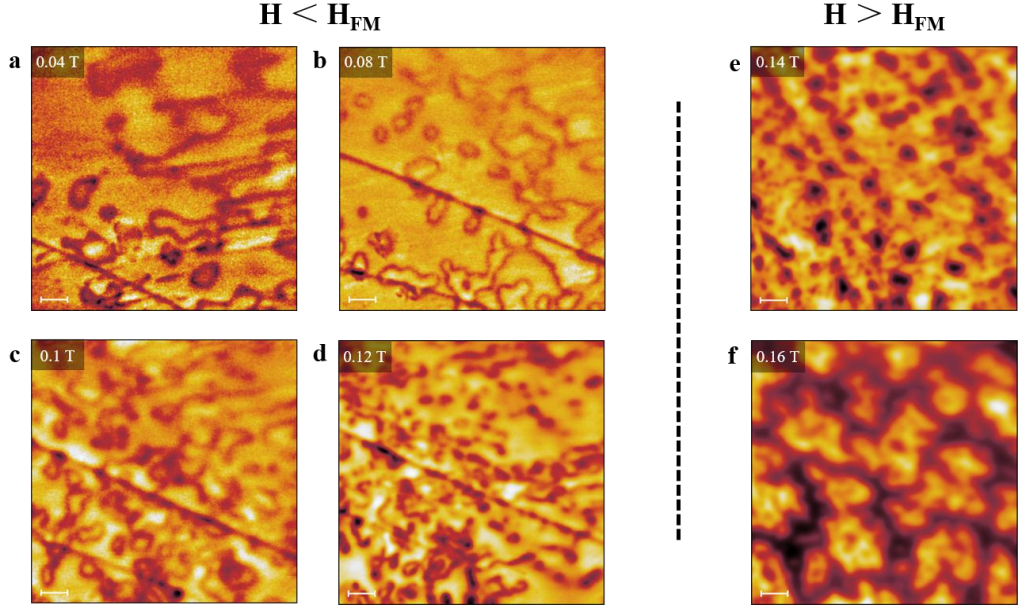

**Fig. 5. MFM images of the field-cooling MnBi$_4$Te$_7$. (a-d)** MFM images taken at 2 K after 0.04, 0.08, 0.1 and 0.12 T field cooling (H < H$_{FM}$). **(e,f)** MFM images taken at 2 K after 0.14 and 0.16 T field-cooling (H > H$_{FM}$). The H$_{FM}$ is the critical field (~ 0.13 T) for the emergence of FM phase in MnBi$_4$Te$_7$ single crystal. Scale bar: 2 μm.

Figs. 5(a-f) shows the MFM images after field cooling through T$_N$ value at different magnetic fields. It is known the AFM-DWs are more energetically favorable in high magnetic fields due to their significant net moment and/or higher magnetic susceptibility. A significant increase of AFM-DW density can be observed at here, since the gain of Zeeman energy reduces the DW energy, and favoring higher DW density at high magnetic fields. Due to the effective reduction of the domain size by field cooling, the significant chiral edge states at the DWs could affect the intrinsic topological surface states.[12] It is also noteworthy that the FM phase appears as the separately distributed circular domains with the magnetic field cooing of 0.14 T. These FM phase domains further expand and connect to separate the AFM phase into small irregular AFM islands with higher magnetic field cooing of 0.16 T. The corrugated MFM contrast of the AFM regions indicate that their flexible magnetic moment orientations are disorderly quenched to a certain extent. How these quenched magnetic phases could affect their topological magnetoelectric properties is still not clear.

**Conclusions**

In summary, our results demonstrate real-space imaging of magnetic domains or DWs at different magnetic phases in AFM-TI MnBi$_4$Te$_7$ single crystal. The presence of multiple magnetic phases provides a prerequisite for the realization of different topological states. Among them, the coexistence of FM and AFM phases is observed in real-space for the first time in an intrinsic AFM-TI. More interestingly, the coexistence of FM and AFM phases induce the metamagnetic behavior of MnBi$_4$Te$_7$ single crystal. Meanwhile, this novel magnetic phase gives us a new platform to explore and realize various topological states. Furthermore, we obtained the complete H-T phase diagram of MnBi$_4$Te$_7$ single crystal by the temperature-dependent study. Finally, using the stray field of the magnetic tip, we successfully achieved the precise manipulation of the surface magnetic structure. By precise magnetic manipulation, we are not only able to eliminate the influence of multiple domains on topological states in TI thin films, but also have the possibility to achieve vertically tunable AFM/FM heterostructures. Our experimental results provide new ideas to realize the control of AFM domains, which can effectively advance the application of AFM spintronic technologies.

**Acknowledgments**

This project is supported by the Ministry of Science and Technology (MOST) of China (No. 2016YFA0200700), the National Natural Science Foundation of China (NSFC) (No. 61674045, 61911540074), the Strategic Priority Research Program (Chinese Academy of Sciences, CAS) (No. XDB30000000). Huan Wang is supported by the Outstanding Innovative Talents Cultivation Funded Programs 2020 of Renmin University of China. Z. H. Cheng was supported by the Fundamental Research Funds for the Central Universities and the Research Funds of Renmin University of China (No. 21XNLG27).

## Materials and Methods

### Sample preparation.

The MnBi$_4$Te$_7$ single crystal were grown by flux method.[35] Mn powder, Bi lump and Te lump were weighed with the ratio of Mn:Bi:Te=1:8:13 (MnTe:Bi$_2$Te$_3$=1:4). The mixtures were loaded into a corundum crucible which was sealed into a quartz tube. Then the tube was put into a furnace and heated to 1100 °C for 20 h to allow sufficient homogenization. After a quick cooling to 600 °C at 5 °C/h, the mixtures were slowly cooled down to 585 °C at 0.5 °C/h and kept for 2 days. Finally, the single crystals were obtained after centrifuging. The plate-like MnBi$_4$Te$_7$ single crystal with centimeter scale can be easily exfoliated.

### EFM and KPFM measurement.

The EFM and KPFM imaging was performed under ambient conditions using an Asylum Cypher S AFM (Oxford Instruments-Asylum Research, Santa Barbara, USA). All EFM (KPFM) data were recorded with a commercial electrostatic tip (Nanosensors, PPP-EFM) in two-pass mode, whereby for each scanned line topography is acquired during the first pass, while the EFM phase (KPFM potential) signal of that line is acquired during the second pass at a set lift height 20 nm.

### MFM measurement.

The MFM experiments were captured by a commercial magnetic force microscope (attoAFM I, attocube) using commercial magnetic tip (Nanosensors, PPP-MFMR-20) based on a closed-cycle helium cryostat (attoDRY2100, attocube). The scanning probe system was operated at the resonance frequency of the magnetic tips, which was around 75 kHz. MFM images were taken in a constant height mode with the scanning plane nominally ~100 nm above the sample surface. The MFM signal, the change of cantilever phase, is proportional to out-of-plane stray field gradient. Dark(bright) regions in MFM images represent attractive (repulsive) magnetization, where magnetization is parallel (anti-parallel) with magnetic tip moments. Finally, AFM domain manipulation was attained using tapping mode.

### XRD and transport measurements.

The single crystal x-ray diffraction patterns were collected from a Bruker D8 Advance x-ray diffractometer using Cu $K_\alpha$ radiation. The measurements of magnetic property were performed on a Quantum Design magnetic property measurement system (MPMS-3).